# CPU Simulation Using Two-Phase Stratified Sampling

Magnus Ekman
NVIDIA
mekman@nvidia.com

*Abstract*—Simulation remains a cornerstone of computer architecture research, yet full end-to-end application execution is prohibitively time-consuming. The industry-standard solution, SimPoint, mitigates this cost by selecting a small number of representative code regions that capture program phase behavior. In this work, we take a fresh look at phase behavior in the SPEC CPU 2017 Integer suite to assess how pronounced such behavior truly is and what accuracy can be expected from typical SimPoint usage. Based on previously published data, we argue that common SimPoint counts can induce substantial estimation errors. To explore this further, we recast SimPoint as a stratified sampling problem, which enables the derivation of a conservative confidence interval. The analysis indicates that significant errors are expected, and our empirical analysis confirms this: with 20 SimPoints, two applications exhibit 40–60% performance prediction error.

We decompose SimPoint into its two fundamental components—stratification (clustering) and sample-unit selection (centroid choice)—and analyze their individual effects on accuracy. We then extend the approach into a two-phase (double) sampling scheme, in which a large preliminary random sample enables improved stratification and more representative region selection. Using this method, the maximum per-application error drops to 3%. Finally, we demonstrate that the proposed two-phase stratified framework achieves an order-of-magnitude reduction in required sample size compared to simple random sampling while maintaining a tight analytical confidence interval, suggesting a practical path toward statistically grounded and efficient architectural simulation.

## I. INTRODUCTION

Simulating applications end-to-end in computer architecture studies is prohibitively expensive in terms of simulation time. Consequently, such studies typically simulate selected regions of the application. SimPoint [1][2] is a widely used technique, both in academia and industry, to identify representative regions by detecting program phases. The process operates as follows: the application is first executed on an ISA-level simulator (as opposed to a cycle-accurate performance simulator) to collect basic block vectors (BBVs) across the full run. The speed of ISA-level simulation makes it feasible to cover the entire execution. The BBVs are then reduced in dimensionality using random projection, after which k-means clustering groups them into clusters, each representing an execution phase. One representative region—a SimPoint, defined as the basic block vector nearest each cluster centroid—is selected per phase.

It is well known [1][2][3] that SimPoint-based performance estimates can incur substantial errors for individual applications, and the technique itself offers no mechanism for detecting or quantifying these errors. A subsequent study [4] showed that accuracy can improve by substantially increasing the number of SimPoints (up to 300 in some cases) and proposed an empirical confidence interval for the baseline configuration. In our experience, simulating hundreds of SimPoints is often impractical due to the warm-up overhead required to avoid cold-start effects for each region (e.g., to warm caches, prefetchers, and branch predictors). In practice, architects often limit themselves to 10–20 SimPoints per application, as evidenced in recent publications [5][6][7]. Based on prior data [3], this can yield performance projection errors in the double-digit percentage range—on the same order as the typical performance improvement between two CPU generations. When comparing projections to current silicon data, such inaccuracies can produce misleading conclusions: a projected next-generation CPU might appear slower than its predecessor, or conversely, performance gains might be overstated. Even when comparing two simulated configurations, seemingly minor system changes, such as varying memory page size, can perturb SimPoint selection sufficiently to render cross-configuration comparisons unreliable.

To address these limitations, we propose a two-phase sampling methodology that both improves projection accuracy and provides a process to quantify estimation error. The first phase uses a large simple random sample to obtain a highly accurate reference estimate. Information from this phase is then used in the second phase to identify more representative simulation intervals than the SimPoint methodology achieves. The performance estimate from the second phase can be directly compared against the first-phase baseline to assess error. Moreover, information from the first phase enables the use of stratified random sampling [8] (Ch. 5) to construct statistically sound confidence intervals. Our study makes the following key contributions:

1. **Two-phase sampling methodology:** We propose and evaluate a two-phase sampling scheme [8] (Ch. 12) in which the first phase is a large simple random sample and the second phase is a stratified sample with one sampling unit per stratum. We show empirically that this approach reduces the maximum observed per-application error from 60% to 3% for individual SPEC CPU 2017 Integer applications, while providing a straightforward mechanism to quantify estimation error.
2. **Confidence intervals:** We demonstrate how to provide tight confidence intervals by periodically performing

simulations with multiple sampling units per stratum. This approach reduces the number of required simulations by an order of magnitude compared to simple random sampling, with only a modest loss in accuracy. We also explain why these efficiency gains are achievable in contrast to prior work [9], which found them difficult to realize in practice.

3. **Analytical treatment of SimPoint as a sampling problem:** We analyze SimPoint through the lens of sampling theory. By applying the collapsed strata method [8] (Sec. 5A.12), we show that it is possible to compute an approximate confidence interval even for the original SimPoint methodology—a capability not previously demonstrated.
4. **Reexamination of phase behavior:** We revisit program phase behavior by plotting CPI distributions for each benchmark. This reveals irregularities, such as outliers, often overlooked by current methodologies.

The rest of this paper is organized as follows. Section II revisits program phase behavior. Section III introduces the two-phase sampling scheme. Section IV describes our experimental methodology, followed by results in Section V. Section VI discusses our findings, Section VII relates them to prior work, and Section VIII concludes. The statistical methods are detailed in Appendix A.

## II. REVISITING PROGRAM PHASE BEHAVIOR

In their seminal work [1], Sherwood et al. demonstrated that computer applications exhibit stable execution phases and showed how this behavior can be exploited to project overall performance as a weighted mean of one simulation point per phase. This methodology—SimPoint—quickly became the industry standard and has served the architecture community well for over two decades. However, many practitioners have observed that program behavior in practice often appears far more irregular than this model suggests, creating a certain cognitive dissonance between theory and experience.

To illustrate this, Fig. 1 plots the distribution of cycles per instruction (CPI[1]) for a large number of randomly selected *simulation regions* [2] in three SPEC CPU 2017 Integer applications. For each application, three charts are shown, corresponding to region lengths of 1 M, 10 M, and 100 M instructions. If applications truly exhibited a small number of well-defined phases with stable CPI, we would expect the charts to show a few sharp spikes, each corresponding to a phase.

Several patterns emerge. First, perlbench and omnetpp show relatively stable phase behavior, whereas xz exhibits a more chaotic distribution spanning a wide CPI range. Second, longer simulation regions appear more stable: in omnetpp, CPI values are more dispersed for 1 M-instruction regions than for 100 M. The x-axis range is based on observed CPI values. For instance, perlbench (1 M region) includes a few regions with CPI $\approx$ 3.0, though the bar is too small to be visible. For longer regions (10 M and 100 M), the highest CPI drops to $\approx$ 2.0, reinforcing that aggregation over longer intervals smooths variability—a natural statistical effect.

These observations connect directly to the evolution of SimPoint itself. The original study [1] used 100 M-instruction regions and selected up to ten SimPoints. Later work reduced the region size to 10 M [2] and eventually 1 M [4], while allowing up to 300 SimPoints to improve accuracy. As reported in [3], the original approach produced a worst-case error of 14.3% (for gcc) and an overall average error of about 3%.

Two insights follow. First, the absence of distinct spikes even at 100 M-region granularity is consistent with prior findings: [4] already noted IPC variation within clusters, and [3] quantified substantial error when using only ten SimPoints. Second, the reduction in error with 1 M regions and 300 SimPoints is unsurprising—random sampling itself achieves low error at such sample sizes [3].

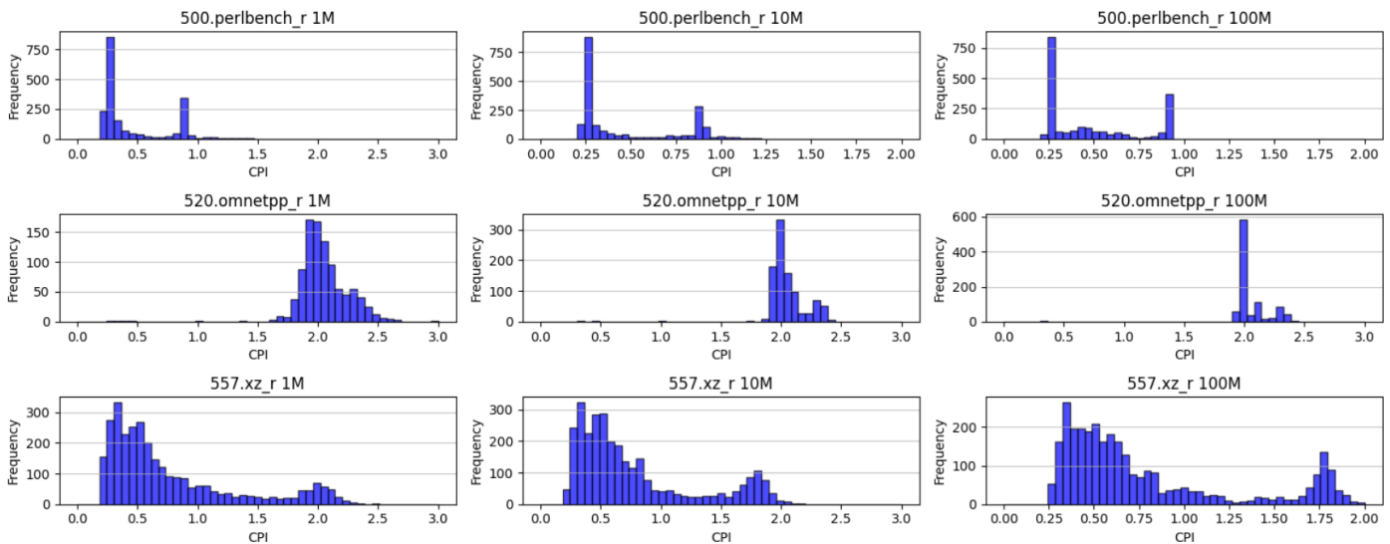

Fig. 1. CPI distributions for perlbench, omnetpp, and xz from SPEC CPU 2017 Int. Each chart is based on many randomly selected regions; region lengths are 1 M (left), 10 M (middle), and 100 M (right) instructions.

---

[1] We use CPI as opposed to IPC so the overall application metric can be computed as an arithmetic mean rather than a harmonic mean.

[2] We use "simulation region" to denote any selected point in the application, not necessarily one chosen using the SimPoint methodology.

To further illustrate, we identified phases using k-means clustering on BBVs following the SimPoint procedure. Fig. 2 shows the resulting CPI distribution for omnetpp with ten clusters and 10 M-instruction regions; bars are color-coded by cluster. While CPI clearly correlates with cluster identity, considerable variation remains within each cluster. This within-cluster variability directly contributes to the inaccuracy observed when using only a limited number of SimPoints.

At this point, one might ask: if this has been understood for two decades, why revisit it? The answer leads directly to our problem statement.

*A. Problem statement*

Simulating hundreds of SimPoints is typically impractical because each simulation must avoid cold-start effects by warming large stateful microarchitectural structures such as caches, prefetchers, and branch predictors. In practice, architects often limit themselves to 10–20 SimPoints per application, each simulated for 10–30 M instructions after warm-up—a practice consistent with several recent publications [5][6][7].

Under these conditions, the expected projection error for individual applications can be 10–15%, comparable to the performance gain of a full CPU generation. This level of uncertainty undermines confidence in pre-silicon projections and makes it difficult to separate sampling error from modeling error when comparing simulation-based results with measurements from real silicon. Reusing identical SimPoints across processor generations can partially mitigate this issue, since relative CPI differences tend to be more accurate than absolute values [10][11][12]. Nevertheless, comparing identical SimPoints is not always feasible, and reducing absolute projection error remains a key objective.

We approach simulation-region selection as a statistical sampling problem. This framing enables the application of established sampling theory to both quantify and reduce projection error.

III. IMPROVING OVER SIMPOINT WITH TWO-PHASE SAMPLING

We begin by revisiting how SimPoint relates to statistical sampling (see Appendix A for concepts and terminology). The key steps of the SimPoint technique are illustrated in Fig. 3. The application is first functionally simulated to collect BBVs. These BBVs are reduced in dimensionality using random projection, followed by k-means clustering[3]. One region—a SimPoint—is chosen per cluster, based on the proximity between the BBV and the cluster centroid.

From a sampling standpoint, SimPoint corresponds closely to stratified random sampling (see Appendix A, Section B). The functional simulation that generates BBVs acts as a census, identifying a stratification variable for each population unit. Clustering forms the strata, and the final SimPoint selection represents sampling one unit per stratum. The key difference is that the selection is *not random* but deterministic—based on proximity to the centroid—so standard analytical formulas for confidence intervals do not directly apply. Nevertheless, the deterministic choice is arguably *better than random*, suggesting that confidence intervals derived for stratified random sampling can serve as conservative bounds for SimPoint. Because only one sampling unit per stratum is used, confidence-interval estimation is non-trivial, but established methods exist to handle this case (see Appendix A, Section C).

*A. Limitations of BBV-based stratification*

A fundamental limitation of the above approach is that CPI does not always correlate well with BBVs. For instance, a function's CPI may vary widely depending on its input data, even if the same basic blocks are executed. Stratifying on a variable that correlates more directly with CPI would yield more accurate projections. The ideal—but impractical—variable would be CPI itself, since measuring CPI for every region

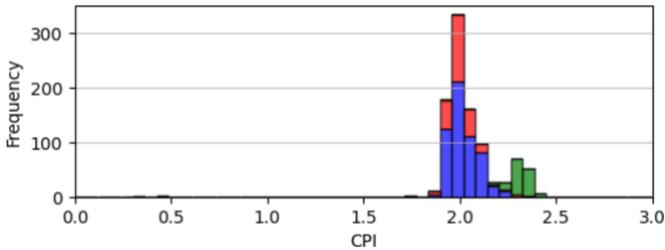

Fig. 2. CPI distribution for omnetpp with color-coded clusters produced by k-means on BBVs.

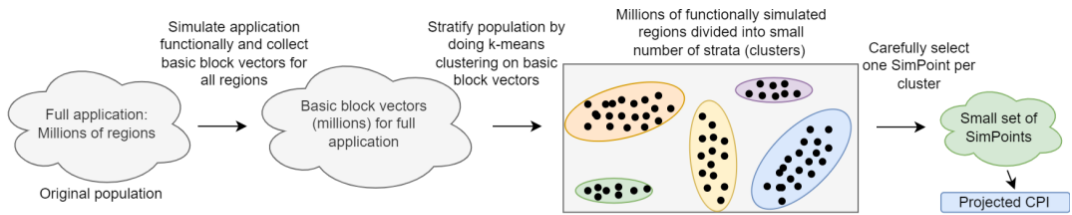

Fig. 3. SimPoint flow. Functionally simulate the full application to produce basic-block vectors (BBVs)—analogous to a census in sampling terms. Reduce BBV dimensionality via random projection, apply k-means clustering (equivalent to stratification), and select one SimPoint—the region closest to each cluster centroid—from each cluster.

---

[3] In this paper, when we talk about clustering BBVs, we always refer to the dimensionality-reduced BBVs.

requires a full population census (detailed simulation of the full application).

As noted in Cochran [8] (Sec. 5A.7), one remedy is to stratify using a historical value of the study variable—for example, from a previous census—and then reuse this stratification in subsequent studies. In our context, this is analogous to simulating the full application once on a baseline (historical) simulator to obtain CPI for every region, then using these values to stratify the regions for future studies on alternative (future) configurations. The assumption is that CPI under the baseline correlates strongly with CPI under modified designs. However, fully simulating even the baseline configuration is often impractical. Instead, we adopt the two-phase sampling approach described in Appendix A, Section D.

### B. Two-phase sampling for simulation

The two-phase scheme begins with a large preliminary random sample, followed by a smaller stratified sample. Applied to simulation, this process is illustrated in Fig. 4.

In the first phase, we randomly select and simulate many regions once using the baseline configuration on a cycle-accurate simulator. This yields detailed performance metrics for each region and a highly accurate estimate of the application's true CPI. The CPI values from this baseline configuration, can then be used to stratify the population for subsequent experiments[4].

Stratifying solely on CPI may, however, conflate regions with different performance causes—e.g., one dominated by cache misses and another by branch mispredictions could share the same CPI. To better distinguish such cases, we propose forming a *Rich Feature Vector (RFV)* comprising CPI together with other microarchitectural metrics such as cache-miss rate and branch-mispredict rate. We then perform k-means clustering on these RFVs to form strata. In the second phase, one or more regions are sampled from each stratum for simulation under new configurations.

Information collected through this process allows us to compute confidence intervals using formulas for two-phase and stratified random sampling[5]. In Section V.A, we evaluate how three stratification schemes—BBV-based, RFV-based, and CPI-only—affect projection accuracy both analytically and empirically.

The region-selection policy (random versus centroid-based) is analyzed separately in Section V.B, leveraging the highly accurate first-phase estimate to quantify the error induced by simulating only a small subset of regions.

### C. Relationship between sampling and program phase behavior

Returning to the charts in Fig. 1 and Fig. 2, one might find them concerning if expecting distinct, constant-CPI phases. Yet from a sampling perspective, such variability poses no conceptual problem. High variance *within* a stratum simply makes for a looser confidence interval or, equivalently, increases the sample size needed to achieve a desired precision.

This interpretation bridges the intuitive "phase behavior" model with the formal framework of stratified sampling, providing a statistically grounded explanation for the performance-prediction variability observed in SimPoint.

## IV. EXPERIMENTAL METHODOLOGY

To quantify the accuracy of the different sampling schemes, we conducted cycle-accurate simulations of a single CPU core implementing the ARMv9 instruction set architecture (ISA). We defined a baseline configuration (Config 0) and six progressively faster alternatives (Configs 1–6).

The baseline is a four-wide out-of-order core with modest cache sizes, a basic stream prefetcher, and a TAGE branch predictor [13]. Each successive configuration introduces incremental microarchitectural enhancements. Table I summarizes the key parameters. Highlighted entries indicate modifications relative to the baseline. The changes include expanded branch predictor tables, larger reorder buffer, wider retire width, reduced memory and cache latencies, and the introduction of two advanced prefetchers: a Spatial Memory Streaming (SMS) prefetcher [14] and a Best Offset (BO) prefetcher [15].

### A. Benchmark suite and sampling setup

We used the SPEC CPU 2017 Integer Rate benchmark suite, compiled with GCC 12. For each configuration, we simulated a large number of 1-million-instruction regions. The rationale for this short region length is that any sampling technique accurate at this granularity will maintain or improve accuracy for longer regions (e.g., 10 M or 100 M instructions), since variance decreases with increased region length.

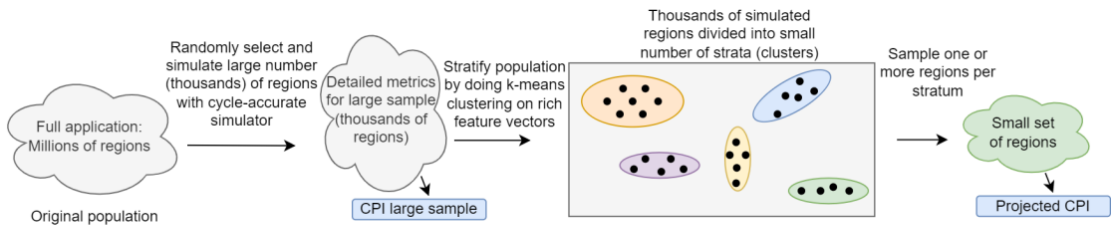

Fig. 4. Conceptual two-phase simulation flow. First, take a large random sample (simulate many regions) to characterize the population and form strata. Then select a small stratified sample for future experiments.

---

[4] Stratifying on baseline CPI might appear as performing data snooping and yield optimistic confidence intervals. This would be the case if estimating the baseline itself, but our goal is to apply the stratification when sampling other configurations. Later sections quantify the resulting error empirically.

[5] In two-phase sampling, both phases contribute to the variance. When the first-phase sample is much larger than the second, its contribution is negligible, and standard stratified-sampling formulas suffice.

TABLE I. PARAMETERS FOR THE SIMULATED CONFIGURATIONS.

|  | Config 0 | Config 1 | Config 2 | Config 3 | Config 4 | Config 5 | Config 6 |
|---|---|---|---|---|---|---|---|
| Fetch width | 8 | 8 | 8 | 8 | 8 | 8 | 8 |
| Issue width | 8 | 8 | 8 | 8 | 8 | 8 | 8 |
| D-cache hit latency | 3 | 3 | 3 | 3 | 3 | 3 | 3 |
| L2 hit latency | 8 | 8 | 8 | 8 | 8 | 8 | 8 |
| I-cache size | 32KB | 64KB | 64KB | 64KB | 64KB | 64KB | 64KB |
| D-cache size | 32KB | 64KB | 64KB | 64KB | 64KB | 64KB | 64KB |
| L2 cache size | 512KB | 1M | 1M | 1M | 1M | 1M | 1M |
| L3 cache size | 2MB | 4MB | 4MB | 4MB | 4MB | 4MB | 4MB |
| SMS PF | Disabled | Disabled | Enabled | Enabled | Enabled | Enabled | Enabled |
| Reorder buffer size | 128 | 128 | 128 | 256 | 256 | 256 | 256 |
| Physical regs | 128 | 128 | 128 | 256 | 256 | 256 | 256 |
| Retire width | 4 | 4 | 4 | 8 | 8 | 8 | 8 |
| Memory latency | 130ns | 130ns | 130ns | 130ns | 90ns | 90ns | 90ns |
| L3 hit latency | 30ns | 30ns | 30ns | 30ns | 20ns | 20ns | 20ns |
| L2 best-offset PF | Disabled | Disabled | Disabled | Disabled | Disabled | Enabled | Enabled |
| TAGE BPU tables | 4 | 4 | 4 | 4 | 4 | 4 | 8 |
| TAGE table entries | 2048 | 2048 | 2048 | 2048 | 2048 | 2048 | 4096 |

Each benchmark was simulated across roughly [6] 1 000 regions, or more when required to achieve a very tight confidence interval under random sampling. The mean CPI computed from this large set was treated as the true reference value against which stratified and two-phase sampling estimates were later evaluated. Table II lists the number of simulated regions per benchmark.

TABLE III. NUMBER OF SIMULATED REGIONS PER SPEC CPU 2017 INTEGER APPLICATION.

| Application | Simulation regions (sample size) |
|---|---|
| 500.perlbench_r | 1,997 |
| 502.gcc_r | 6,195 |
| 505.mcf_r | 964 |
| 520.omnetpp_r | 967 |
| 523.xalancbmk_r | 6,861 |
| 525.x264_r | 915 |
| 531.deepsjeng_r | 1,041 |
| 541.leela_r | 1,062 |
| 548.exchange2_r | 1,030 |
| 557.xz_r | 3,047 |

*B. Rich feature vector metrics*

For each simulated region, we recorded 38 microarchitectural performance statistics grouped into the categories shown in Table III. These metrics were used to form the Rich Feature Vector (RFV) used for stratification (see Section III). Given the modest number of metrics, we didn't reduce dimensionality with random projection, but we did standardize the values.

TABLE II. PERFORMANCE COUNTER EVENTS COLLECTED FROM THE SIMULATOR.

| Category | Metric |
|---|---|
| Global | Cycles per instruction |
| Frontend events | Branch mispredicts, Cond. branch mispredicts, Target branch mispredicts, i-cache misses, i-TLB misses |
| LSU events | L1d access, L1d load miss, L1d store miss, L1d total miss, L1d writeback |
| L2 cache events | L2 cache misses, L2 cache load misses, L2 cache writebacks |
| L3 cache events | L3 cache read accesses, L3 cache write accesses, L3 cache misses |
| Top-down stack stalls | 21 different stall bins spanning frontend and backend of the core |

## V. EXPERIMENTAL RESULTS

Fig. 5 reports per-application IPC for each simulator configuration using the sample sizes in Table II. The 95%

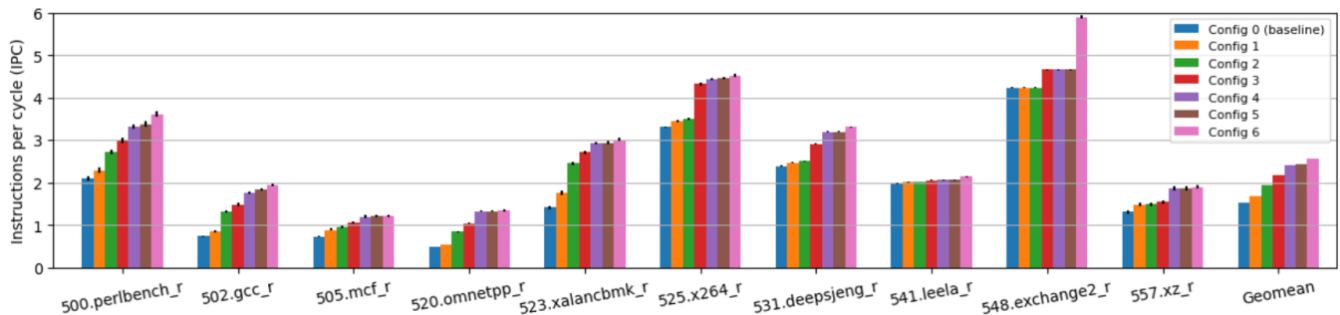

Fig. 5. IPC and associated confidence intervals for each application across all seven configurations; intervals computed with the sample sizes in Table II.

---

[6] The exact number of collected regions varied due to implementation details in our infrastructure.

confidence intervals are extremely tight (the small error bars atop each bar). The overall geomean IPC ranges from 1.52 (Config 0) to 2.56 (Config 6), i.e., +68% over baseline—representing a multi-generational performance spread. We deliberately used such a wide range to cover both incremental and disruptive design studies. As a by-product of the large samples, we were able to visualize CPI distributions per application for the baseline configuration (Fig. 6).

## A. Stratification schemes with one unit per stratum

We used the simulation data above to evaluate several sampling schemes and compare their analytical confidence intervals. For context, we included simple random sampling. The three stratified schemes differed only in stratification:

- **BBV:** k-means clustering on instruction-level BBVs (SimPoint-style).
- **RFV:** k-means clustering on baseline cycle-accurate metrics (Table III)—e.g., cache-miss and branch-mispredict rates—including CPI.
- **Dalenius–Gurney:** strata formed solely on baseline CPI (Appendix A, Section E), with stratum sizes chosen to reduce variance.

To keep comparisons simple, we fixed sample size = 20 for all schemes and compared the resulting confidence intervals. The chosen sample size is similar to what is often used in practice [5][6][7] and approaches the size where the central limit theorem is applicable for confidence interval calculations. For the stratified schemes, we used one unit per stratum (20 strata). Although one-unit-per-stratum is an extreme design point, it is an established statistical method [16][17][18][19] and in line with SimPoint. A challenge is that within-stratum variance cannot be directly estimated from one unit per stratum[7]. For experimental purposes, we already had many simulated regions, enabling us to compute within-stratum variance and thus analytical confidence intervals. Section V.A.3) evaluates a practical alternative (collapsed strata) that does not require many sampling units per stratum. For the comparisons in this subsection, we selected the region from each stratum randomly, while Section V.B.1) analyzes the impact of using centroid selection (like SimPoint).

### 1) Theoretical confidence intervals

We stratified using the methods above. Stratifications that rely on cycle-accurate metrics (RFV and Dalenius–Gurney) were built from Config 0 data. We then computed per-stratum variances and the resulting confidence intervals for Config 6—the configuration farthest from baseline—to stress how well baseline-driven stratification carries over to a different design.

Because k-means is stochastic, we repeated clustering ten times with different seeds for the BBV and RFV schemes, yielding 10 points per application. Random and Dalenius–Gurney yielded only one point each. The resulting 95% confidence intervals, stated as a percentage (margin of error) are shown in Fig. 7, indicating the following key results:

- **Random sampling** often has very loose intervals at n = 20: >30% for gcc and xz, and ≈20% for perlbench, mcf, and xalancbmk—as expected for small n.
- **Stratified on BBV** is surprisingly worse than random sampling for about half the applications (gcc, mcf, omnetpp, xalancbmk, xz), with one gcc case exceeding 70% margin of error. (A 70% margin at 95% confidence means that in 5% of repetitions, error can exceed 70%—clearly risky.)
- **Stratified on RFV** and **Dalenius–Gurney** perform much better: most margins are <10%; gcc is ≈20%—still far better than Random.

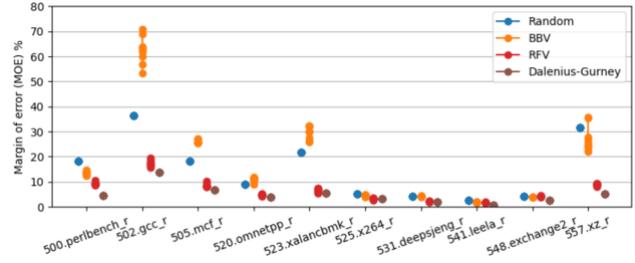
Fig. 7. Analytical confidence intervals for the four sampling schemes.

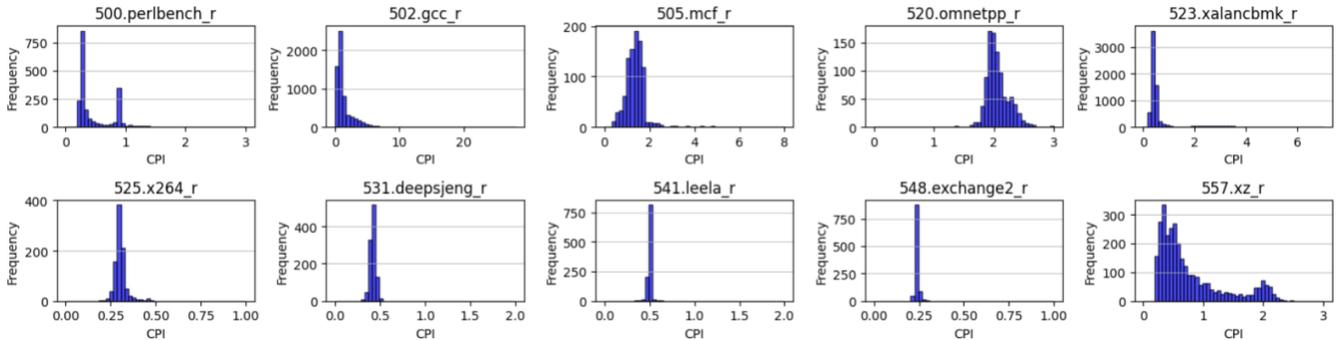
Fig. 6. CPI distributions for the baseline configuration across all applications.

---

[7] At least two sampling units per stratum are required, as described in [8] (Sec. 5.4).

The loose confidence intervals for gcc is explained by its high variance. Its baseline CPI averages 1.36 but includes extreme outliers near 28, >20× the mean. Including one such outlier in 20 data points can nearly double the estimate. The high CPI is caused by an L2-miss chain in the baseline configuration, while for the best configuration, CPI improves to 5.66 thanks to SMS prefetching and reduced memory latencies.

To further understand why stratified random sampling on BBV performed worse than simple random sampling for gcc, we evaluated #strata = 50, 500, 1,000. The margin of error fell to 35%, 12.5%, and 10.5%, respectively, but random also improved with larger n, so sampling efficiency remained higher for random than for BBV at these sizes. Section V.D revisits this with >1 unit per stratum, leading to a different result.

*2) Empirical confidence intervals*

Analytical intervals rely on the central limit theorem, assuming large sample size. A sample size of 20 is on the borderline for this assumption to be valid. We therefore ran Monte Carlo experiments: for a fixed clustering, we repeatedly drew random samples (1,000 trials) and recorded the 5th-percentile worst-case errors (minimum error among the 50 worst outliers). Fig. 8 summarizes the results. Empirical margins broadly agree with the analytical ones in Fig. 7, supporting the viability of computing intervals with 20 strata × 1 unit, provided per-stratum variances are known.

*3) Practically computable confidence interval*

In practice, we don't have many samples per stratum (nor 1,000 Monte Carlo trials). To estimate variance with one unit per stratum, we used the method of collapsed strata (Appendix A, Section C): pair strata judged to be similar and estimate within-stratum variance from the two observations in each pair. We did this grouping by first ordering the strata based on CPI for Config 0 and then forming pairs from neighboring strata. This yields approximate intervals. Fig 9. shows the resulting margins of error.

Notably, the chart uses only 20 simulations (one per stratum) for Config 6. This implies a practical way to compute confidence intervals for SimPoint-like designs: treat deterministic within-stratum selection (e.g., centroid selection) as "better than random" and use collapsed-strata confidence intervals. Some bars deviate from Fig. 7 and Fig. 8, which is expected; collapsed-strata variance estimates can be unstable [16][19].

*B. Impact of selection scheme*

So far, intervals are often too loose for practical use—especially for BBV-based stratification (up to ~50%), and even RFV stratifications can be 15–20% in tough cases. The remedy is to abandon random within-stratum selection. We evaluated two different schemes, starting with centroid-selection.

*1) Centroid-selection*

SimPoint selects, per cluster, the region whose BBV is closest to the centroid. We analogously applied centroid selection to RFV clusters and to Dalenius–Gurney strata (where "centroid" reduces to the mean CPI of the stratum). Fig. 10 reports measured errors (not confidence intervals): each vertical line shows errors for Configs 0–6 for a given app/scheme. Each individual marker corresponds to one of the seven simulator configurations. The results are striking: RFV and Dalenius–Gurney yield low single-digit errors in most cases. BBV still shows large outliers for some apps, which is expected when using a small set of clusters [1][3][4]. We explored this further by increasing the number of clusters from 20 → 50 for gcc, which reduced max error to 5.4%. The key takeaway is that BBV can be accurate if using many regions but stratifying on performance-related features (RFV) requires fewer regions than BBV to reach good accuracy.

*2) Mean-selection*

To separate the effect of stratification from selection policy, we did an experiment where we kept the same strata but, instead of centroid selection, chose the region whose baseline CPI (Config 0) was closest to the stratum mean CPI ("mean selection"). The reasoning was that if a given simulation region CPI is close to the stratum mean CPI for Config 0, it is also likely to be close to its stratum mean for other configurations. Fig. 11 shows the results.

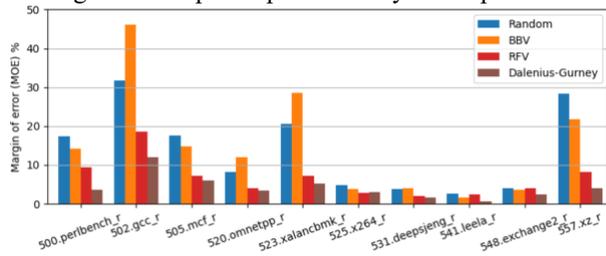

Fig. 8. Empirical confidence intervals for the four schemes

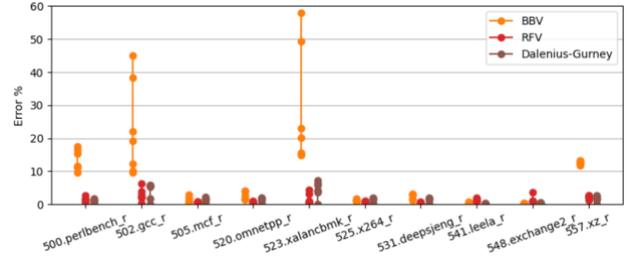

Fig. 10. Errors from centroid-selection for the three stratification schemes.

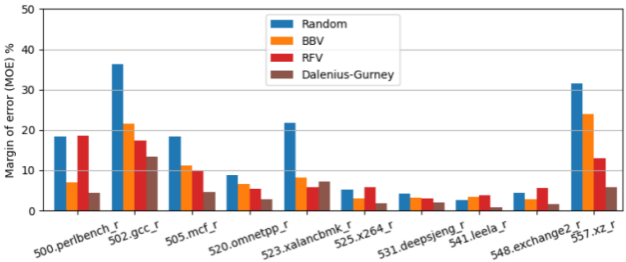

Fig. 9. Confidence intervals computed using collapsed strata.

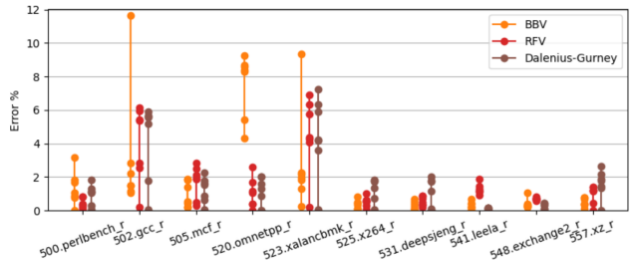

Fig. 11. Errors from mean selection for the three stratification schemes.

Mean selection greatly improves BBV, removing the worst outliers (note the reduced y-axis vs. Fig. 10). Still, BBV remains inferior to RFV and Dalenius–Gurney. Including metrics directly related to performance results in a better stratification than stratifying solely on BBVs. The results from comparing RFV and Dalenius–Gurney are mixed. Still, intuitively RFV is safer because equal CPI can arise from different causes (e.g., cache misses vs. branch mispredictions). Similarly, centroid selection can be preferable to mean selection under RFV, since the centroid leverages the entire feature vector.

*C. Resulting distributions*

Accurately estimating the mean does not guarantee that selected regions span the application's behavioral diversity. Fig. 12 shows the distributions approximated by the 20 selected regions for the baseline configuration using RFV stratification with centroid selection. The data points used for each chart were synthesized using weights and CPI for the 20 regions. The number of data points matches the sample sizes from Table II, making the charts directly comparable to the distributions from Fig. 6. Compared to Fig. 6, some nuances are missing—e.g., xz has no points between 1.5–2.0 CPI; the CPI ranges for xalancbmk and leela are reduced.

With 500 strata (Fig. 13), the approximations resemble baseline distributions more closely, but leela's CPI range still doesn't match Fig. 6. Such blind spots can be addressed with targeted runs. Even if they have negligible impact on the mean, they still represent aspects of application behavior.

*D. Two-phase sampling with multiple units per stratum*

Section V showed empirically that errors remain low across substantial microarchitectural changes, but this is not guaranteed in all cases. A practical safeguard is to periodically run a larger sample. Rather than re-running the full initial random set, we can leverage existing stratification: increase units per stratum to obtain tight confidence intervals without returning to the original sample size. Use this estimate to validate that the smaller set of selected regions are still representative.

This produces a two-phase design: Phase 1 is the large simple random sample; Phase 2 is a smaller stratified random sample. The total variance (and confidence interval width) consists of a combination of variance introduced from both phases. In Section V.A, Phase-1 variance was negligible in comparison to Phase-2 variance, which was high due to sampling a single unit per stratum. With an increased Phase-2 sample size, both terms matter and must be included (see (5) and (6) in Appendix A, Section D).

To evaluate the efficiency of the two-phase sampling scheme, we set the following policy: tolerate at most a 50% increase in margin of error relative to the original random sample. For example, if the Phase-1 margin is 2%, size the Phase-2 stratified sample so the combined margin is ≤3%. In other words, we selected the stratified sample size such that the error contribution from the original random sample became the dominating factor. Table IV lists the resulting sizes and margins.

The initial random sample required 24,079 simulations. With 20 strata, day-to-day studies can use 20 regions per application (200 total for the benchmark suite) — a ~100× reduction with only a small accuracy loss. Periodic multi-unit stratified runs (RFV-based) to quantify the accuracy require 1,917 simulations—~10× fewer than the original 24,079—to meet the target precision. Using BBV instead would require 6,818 simulations to reach the same accuracy.

A final note on the two-phase confidence interval formula (6) in Appendix A, Section D: the Phase-1 sample influences the final confidence interval **(i)** directly via its sample size, and **(ii)** indirectly by shaping stratum weights and within-stratum variances used in Phase 2. The actual CPI values from Phase 1 do not enter the confidence interval formula once stratification is fixed. Thus, after strata are defined, one can compute the confidence interval from Phase-2 data alone—a property with useful practical implications discussed next.

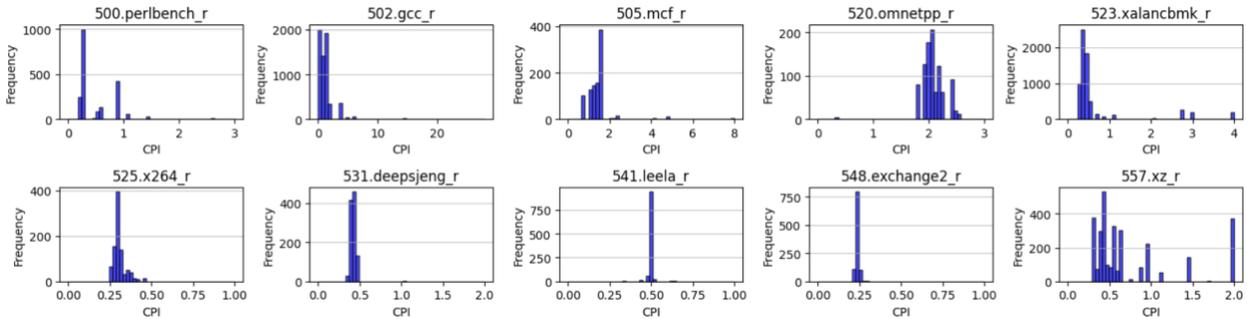

Fig. 12. CPI distributions approximated by a 20-region stratified sample.

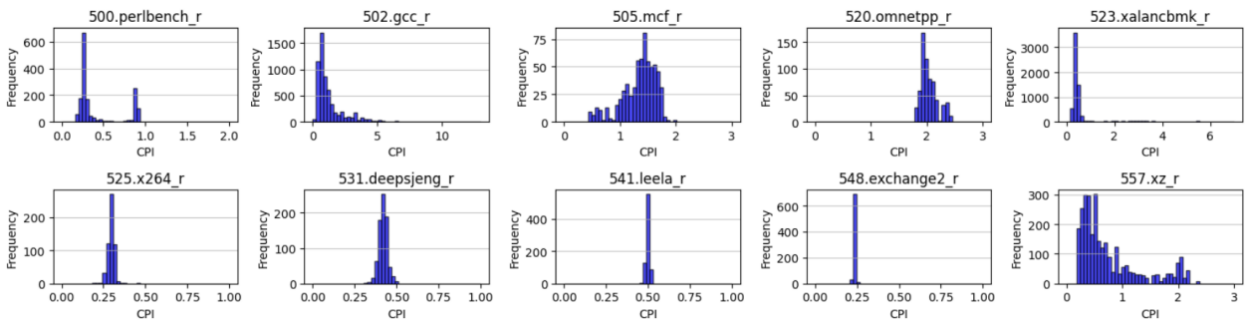

Fig. 13. CPI distributions approximated by a 500-region stratified sample.

TABLE IV. NUMBER OF SIMULATION REGIONS FOR THE TWO-PHASE DESIGN WITH MULTIPLE UNITS PER STRATUM (20 STRATA).

| Application | Regions (random) | Margin of error (random) | Regions (stratified RFV) | Regions (stratified BBV) | Margin of error (stratified) |
|---|---|---|---|---|---|
| 500.perlbench_r | 1,997 | 1.7% | 117 | 286 | 2.6% |
| 502.gcc_r | 6,195 | 1.9% | 561 | 2546 | 2.9% |
| 505.mcf_r | 964 | 2.5% | 90 | 277 | 3.7% |
| 520.omnetpp_r | 967 | 1.2% | 140 | 250 | 1.8% |
| 523.xalancbmk_r | 6,861 | 1.1% | 307 | 1944 | 1.6% |
| 525.x264_r | 915 | 0.72% | 140 | 198 | 1.1% |
| 531.deepsjeng_r | 1,041 | 0.54% | 147 | 404 | 0.80% |
| 541.leela_r | 1,062 | 0.35% | 82 | 174 | 0.52% |
| 548.exchange2_r | 1,030 | 0.57% | 211 | 345 | 0.85% |
| 557.xz_r | 3,047 | 2.4% | 122 | 394 | 3.5% |
| **Total** | **24,079** | **N/A** | **1,917** | **6,818** | **N/A** |

## VI. DISCUSSION

Building on the results in Section V, we now present a recommended methodology, highlight key observations, and outline improvement opportunities to research further.

### A. Recommended methodology

Fig. 14 summarizes our recommended flow. With an additional up-front cost, it delivers higher accuracy than SimPoint for a given sample size and, crucially, quantifies the resulting error. The methodology consists of the following steps:

1. **Initial characterization.** Simulate a large number of random simulation regions once to obtain an accurate estimate of overall CPI. In our evaluation, we directly used the sample sizes in Table II, but in practice you can iterate: start small, estimate variance, then scale to meet a target confidence.

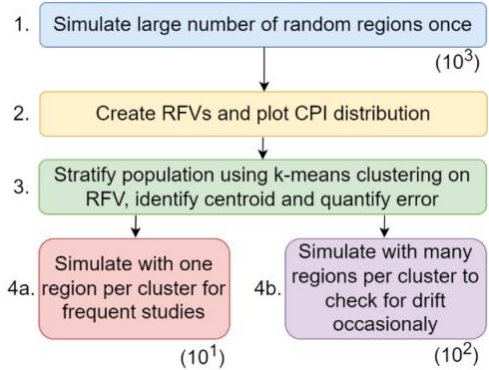

Fig. 14. Recommended methodology. Numbers in the lower-right corners indicate typical simulations per application for that step.

2. **Construct RFVs and visualize distributions.** Form rich feature vectors (RFVs) from the initial runs and plot CPI distributions (as in Fig. 6). These plots are more informative than expected and help decide the number of clusters. Heavily dispersed apps (e.g., gcc, xz) benefit from more clusters than relatively stable ones (e.g., leela).

3. **Stratify via k-means on RFVs; pick centroids.** Cluster RFVs with k-means, then select the region closest to each cluster centroid. Compare the resulting estimate to the accurate baseline from Step 1 (as in Section V.B.1)). This error is not representative of what will be seen for other configurations, yet Fig. 10 shows it remains low in practice across large design changes. Choosing k: we used $k = 20$ for simplicity (aside from sensitivity studies for gcc). Alternatives include Bayesian Information Criterion (BIC) as in [1] or heuristics guided by the CPI distribution from Step 2[8].

4. **Periodically perform confidence interval checks.** Use one region per cluster for day-to-day studies (4a in Fig. 14), i.e., a SimPoint-like flow but with higher accuracy. Occasionally, or when drift is suspected, run multiple regions per cluster (4b in Fig. 14) to obtain tight confidence intervals (Section V.D). If CPI(4a) ≈ CPI(4b), continue using the same regions going forward.

### B. Observations

We make the following observations:

- **Orders of magnitude.** The order of magnitude of simulation counts per application in Fig. 14 are $10^3$ for initial characterization, $10^2$ for occasional confidence interval tightening, and $10^1$ for frequent day-to-day studies. Some applications exhibit sufficiently low variance that $10^2$ may suffice for Step 1.

- **Cost perspective.** Characterization (~$10^3$ sims) is ~100× a single day-to-day study (~$10^1$). In industry (e.g., ~30 engineers, multiple studies/day), this can be amortized to roughly one day of simulations per app. In academia, a single publication often involves ≥100 simulations; reused regions across papers/students translate to a ~10–20% overhead.

- **Checkpoint storage.** Maintaining many regions implies many checkpoints. The initial $10^3$ need not persist; still, to

---

[8] We caution against using the baseline's small quantified error to justify reducing $k$, as that error is inherently small. A large quantified error, however, clearly suggests increasing $k$.

enable periodic multi-unit strata checks, keeping ~10× more checkpoints is required. Storage can be mitigated with compact checkpoint formats that retain only state touched within the interval [20].

- **Outliers and blind spots.** Some applications exhibit surprising outliers that may be absent from the small selected set. While they rarely affect the mean CPI, they do represent true application behavior. Comparing the 20-region approximations (Fig. 12) with the full distributions (Fig. 6) enables spotting gaps (e.g., xz lacking CPI in 1.5–2.0, or missing leela regions >1.0 CPI), that can be studied separately.
- **Microarchitecture dependence.** Unlike SimPoint, our clustering uses performance-related metrics, raising concerns about portability across designs, especially if simulation region selection is never updated. In practice, microarchitectures evolve incrementally, and simulation regions are re-selected over time as binaries/compilers change, and Section V.B.1) shows robustness even across significant changes. If/when drift occurs, Step 4b detects it.

### C. Future research directions

The experimental results reveal the following future research directions:

- **Leverage silicon measurements.** Evaluate the benefit of using silicon data to estimate initial sample sizes and to create CPI distribution charts even before the flow in Fig. 14.
- **Cheaper characterization with faster simulator.** Steps #1–#3 could use a faster, approximate performance simulator that still yields RFVs. Instead of quantifying error at Step 3, one could then compute confidence intervals at Step 4b using the accurate simulator. As noted in Section V.D, Phase-1 CPI values are not required to compute the Phase-2 confidence intervals once strata are formed.
- **Microarchitecture-independent features.** Alternatively, research how to build strata from ISA-level metrics (e.g., [21]) rather than micro-architecture dependent metrics—conceptually similar to the previous bullet. The Step 4b confidence interval still provides a high-accuracy ground truth and bounds the error of the 4a estimates.
- **Extend to multi-core.** Nothing fundamental prevents applying the method to multi-core simulation but its feasibility needs to be investigated. One could construct system-level RFVs by combining (e.g., averaging or concatenating) per-core metrics. Some multi-core workloads show less distinct phases [11], but stratified sampling does not require low within-stratum variance to be valid (Section III).

## VII. RELATED WORK

This section discusses prior work most relevant to our study. SimPoint is discussed in Section VII.A, prior applications of stratified sampling to computer system simulation in Section VII.B, and other sampling-based simulation techniques in Section VII.C.

### A. Selecting representative regions with SimPoint

The SimPoint methodology, introduced by Sherwood et al. [1], has since been analyzed and extended in several follow-up studies [2][4][10][22]. SimPoint first collects basic block vectors (BBVs) using a functional (ISA-level) simulator, reduces their dimensionality via random projection, and applies k-means clustering to group the vectors into disjoint sets representing execution phases. The simulation region whose BBV is closest to each cluster centroid is selected as the SimPoint for that phase, and the application's overall IPC is estimated as the weighted mean across these points.

The original work used 100 M-instruction intervals and up to ten SimPoints. Two drawbacks were identified: (i) high per-application error (up to 14.3 % for gcc [3]) and (ii) the absence of a formal error bound. Later work [4] increased accuracy by using as many as 300 SimPoints. The technique was also modified to guide the clustering step by statistical analysis and provide a probabilistic error bound—though they noted that this bound is not applicable to other simulator configurations.

Our approach addresses the same problem but improves SimPoint in two key ways:

1. **Lower error with small sample sizes.** We achieve high accuracy without expanding the number of simulation regions into the hundreds.
2. **Practically applicable technique to quantify error.** Our two-phase sampling design provides a statistically grounded confidence interval by first collecting a large random sample for stratification and baseline estimation, and then periodically sampling multiple regions per cluster to recompute confidence intervals using formulas from stratified random sampling.

### B. Stratified sampling for computer system simulation

Stratified sampling has previously been examined for architectural simulation. The earliest study we are aware of [9] identified large theoretical potential (a 43× reduction in required simulations) but achieved only 2.2× in practice when stratifying on BBVs. The authors also deemed k-means impractical for short regions due to the sheer number of BBVs to cluster. Our findings differ for several reasons:

- First, [9] required at least 30 samples per stratum to compute a confidence interval—an overly conservative assumption. Classical sampling theory [23] Ch. 4.2 requires either a large number of strata or a large number of samples per stratum.
- Second, we employ a two-phase sampling approach, enabling richer stratification variables (RFVs rather than BBVs). As shown in Table IV, stratifying on RFVs yields roughly a 12.5× reduction, versus 3.5× for BBVs for the full SPEC 2017 Integer benchmark suite.
- Third, the two-phase design mitigates scalability issues. Instead of clustering on BBVs for the entire application, one can draw a large random sample of BBVs (e.g., ≈100k) and cluster on that sample; the resulting variance from Phase 1 is negligible, effectively reducing the problem to a standard one-phase stratified sample.

Hence, we believe that these items combined remove the barriers that prevented the previous study [9] from reaping benefit from stratified sampling.

Subsequent successful uses of stratified sampling include Program-Structure-Aware Stratified Sampling (PASS) [24], which uses calling context and loops for stratification, achieving a 3× reduction in sample size versus simple random sampling. Another study [25] examined the trade-off between sample size, interval length, and number of phases (strata) when stratifying on BBVs. Finally, SimProf [26] applied stratified random sampling to data-analytics workloads on silicon, stratifying sampling units based on similarity in their call stacks. This reduced the mean-error from 8.9 % → 1.6 % compared to simple random sampling.

*C. Other simulation techniques with sampling*

Several other works apply sampling to architectural simulation. Simple random sampling for estimating mean IPC appears in [3][27], and similar methods quantify differences between configurations rather than absolute performance [11][12]. These approaches target the same fundamental problem as ours but lack the efficiency gains of stratified sampling.

Sampling has also been applied to multi-core simulation [11][22][28], though our focus is single-core. A related study [29] used sampling to address timing noise from race-condition variability in multi-threaded runs—a distinct goal from ours.

VIII. CONCLUSIONS

End-to-end architectural simulation of full applications remains prohibitively time-consuming. For two decades, both academia and industry have relied on SimPoint to address this challenge by exploiting application phase behavior. Yet it has long been recognized that SimPoint can produce substantial estimation errors for individual applications and provides no systematic means to quantify them.

The central contribution of this study is the introduction and evaluation of a two-phase sampling scheme. The first phase uses a large simple random sample, and the second applies stratified sampling with a single sampling unit per stratum. This approach yields higher accuracy than SimPoint and, importantly, provides a straightforward mechanism to quantify estimation error. In our experiments on the SPEC CPU 2017 Integer suite, we observed per-application errors of up to 60% with 20 SimPoints, whereas the proposed scheme reduced this to 3% using the same number of regions.

A second contribution is the demonstration of how to obtain tight confidence intervals by periodically simulating multiple units per stratum. Previous studies have noted the theoretical efficiency of stratified sampling but struggled to realize it in practice. By combining stratification with two-phase sampling, we achieve an order-of-magnitude reduction in required simulations relative to simple random sampling, with only a modest loss in accuracy.

Finally, our study provides a fresh perspective on program phase behavior. By plotting CPI distributions for each benchmark, we reconcile the classical notion of stable phases with the often-chaotic behavior observed in practice. These charts proved surprisingly informative. Comparing such charts with corresponding charts for selected simulation regions revealed omitted outliers that may merit targeted investigation.

APPENDIX A – SAMPLING TECHNIQUES

This appendix briefly describes the sampling techniques used in this paper. For more extensive descriptions see [8] and [23]. As we apply sampling to computer system simulations, our goal is to provide an estimate of the overall CPI by simulating only a subset of an application. In sampling terminology, the application is the sampled population, and the simulated regions are sampling units (individuals). From the simulated regions (the sample), we compute a performance estimate, stated in the form of a confidence interval (CI):

$$CI = \bar{y} \pm z_{\alpha/2}\sqrt{v(\bar{y})}, \quad (1)$$

The first term, $(\bar{y})$, represents the sample mean, and the second term ($z_{\alpha/2}\sqrt{v(\bar{y})}$) represents a margin that, with a certain level of confidence, is expected to include the true population mean.

The desired confidence level (e.g., 95%) determines the value of $z_{\alpha/2}$, which can be obtained from a normal distribution table when the sample size is large (typically $n \geq 30$). With smaller sample sizes, the z-value is replaced by a t-value obtained from the t-distribution. The t-value depends not only on the desired confidence level but also on the degrees of freedom (*df*), which in turn depend on the sample size and sampling scheme.

As shown in the formula above, the width of the confidence interval also depends on the standard deviation[9] of the sample mean, $\sqrt{v(\bar{y})}$. This quantity is distinct from the standard deviation of the population itself: it reflects both the population variability and the effect of the sampling method and sample size. The following sections describe different sampling methods and formulas for computing *df* and $\sqrt{v(\bar{y})}$.

*A. Simple random sampling*

In simple random sampling (SRS), sampling units are selected at random from the population. The estimators for the population mean $\bar{y}$, the population variance $s^2$, and the variance of the sample mean $v(\bar{y})$ are:

$$\bar{y} = \frac{1}{n}\sum_{i=1}^{n} y_i, \; s^2 = \frac{1}{n-1}\sum_{i=1}^{n}(y_i - \bar{y})^2, \; v(\bar{y}) = \frac{s^2}{n} \quad (2)$$

If the sample size is small ($n < 30$) the degrees of freedom $df = (n-1)$ is used to obtain the t-value to compute the confidence interval.

*B. Stratified random sampling*

Stratified random sampling is often more efficient than simple random sampling in that it can lead to higher accuracy for a given sample size. The first step is to divide the population into *L* non-overlapping strata that together cover the entire population. Simple random sampling is then applied separately within each stratum.

The key metrics needed for the confidence interval are estimated as follows:

$$\bar{y} = \sum_{h=1}^{L} W_h \bar{y}_h, \qquad v(\bar{y}) = \sum_{h=1}^{L} \frac{W_h^2 s_h^2}{n_h} \quad (3)$$

Here $\bar{y}_h$ and $s_h$ denote the mean and standard deviation of the sample of size $n_h$ within stratum *h*. The term $W_h$ is the weight of stratum *h*, defined as $W_h = \frac{N_h}{N}$, where $N_h$ is the population size of stratum *h* and $N$ is the total population size.

The variance of the estimated mean, $v(\bar{y})$, depends only on the within-stratum variances. Consequently, if strata are formed so that the units within each stratum are similar, the resulting confidence interval will be narrower. To achieve this property, stratification is typically done based on an auxiliary variable *x* that is expected to correlate well with the study variable $y$. Ideally, the value of $x$ is known for every unit in the population, for example from a previous census[10]. In practice, $x$ can often be a historical value of $y$ obtained in an earlier study, since it usually correlates strongly with the current value of $y$ [8] (Sec. 5A.7).

---
[9] Standard deviation is the square root of the variance.

According to [23] (Sec. 4.2), a z-value may be used when either each stratum sample size is large, or the number of strata $L$ is large. Otherwise, the number of degrees of freedom ($df$) for the t-value can be approximated using Satterthwaite's formula [30]. A simpler rule-of-thumb often used in practice [31] is $df = n - L$.

*C. Stratified sampling with one unit per stratum*

Stratified random sampling can be taken to the extreme by selecting only a single sampling unit from each stratum. A challenge in this case is that the within-stratum variance cannot be estimated directly, since at least two observations per stratum are required to compute a sample variance. To approximate this quantity, the method of collapsed strata can be used [8] (Sec. 5A.12). In this approach, adjacent strata that are expected to be similar are grouped together, and the variance for each collapsed pair is computed as:

$$s_h^2 = s_{h+1}^2 = \frac{(y_h - y_{h+1})^2}{4}, \quad n_h = n_{h+1} = 1 \quad (4)$$

In this formula $y_h$ and $y_{h+1}$ are the sampled values from the two strata being combined. Each collapsed pair provides one estimate of the within-stratum variance that is then applied to both strata. For this method to perform well, it is important to form pairs of strata whose sampled units are expected to have similar values.

When only one unit is selected per stratum, and the number of strata $L$ is small, the recommended degrees of freedom ($df$) for use with the t-distribution is $df = L - J$, where J is the number of collapsed groups [18]. In the case of pairwise collapsing, $J = \frac{L}{2}$, resulting in $df = \frac{L}{2}$.

*D. Two-phase sampling for stratification*

As mentioned in Section B, it is common to stratify the population on a variable $x$, that is known for every unit in the population, and is expected to correlate well with the study variable $y$. When such an auxiliary variable is not available for the entire population, two-phase sampling (also known as double sampling) can be used [8] (Ch. 12). In the first phase, a large simple random sample of size $n$ is drawn to collect values of $x$. Data from this sample is then used to stratify the population, after which a smaller second-phase stratified sample is drawn to measure $y$.

This approach is beneficial when the cost of obtaining $x$ is lower than that of $y$, or when $x$ can be reused across multiple future studies so that the cost of the first phase can be amortized[11].

The overall mean is computed as in stratified sampling, but the variance of the sampled mean has an additional component reflecting the first-phase sample:

$$v(\bar{y}) = \frac{s^2}{n} + \sum_{h=1}^{L} \frac{W_h^2 s_h^2}{n_h} \quad (5)$$

The first term represents the variance from the initial simple random sample and the second term represents the variance from the stratified (second-phase) sample. When the first-phase sample size $n$ is much larger than the combined second-phase sample size (the sum of all $n_h$), the first term becomes negligible, and the variance reduces to that of ordinary stratified random sampling.

The first term in the formula above includes the variance of the overall population as obtained by the first-phase sample. The following alternative formula computes the variance of the sampled mean using data solely from the second-phase stratified sample:

$$v(\bar{y}) = \frac{1}{n}\sum_{h=1}^{L} W_h(\bar{y}_h - \bar{y})^2 + \sum_{h=1}^{L} \frac{W_h^2 s_h^2}{n_h} \quad (6)$$

This allows for doing the first-phase sample only once to stratify the population and later compute accurate confidence intervals for multiple subsequent stratified samples as the population changes.

*E. Stratification using the Dalenius/Gurney method*

The stratification method can significantly impact the efficiency of stratified random sampling. The best possible stratification will be obtained by stratifying on the target variable ($y$) itself, but it should be noted that this will lead to an optimistic confidence interval and should be avoided. However, as previously described, one option is to use a historical value of $y$ as the auxiliary variable $x$ that is used for stratification. Various methods have been proposed to identify optimal stratification. One simple approximate method proposed by Dalenius and Gurney [32] is to order the individuals based on $x$, and then select strata boundaries such that the product of weight and standard deviation for all strata are identical:

$$W_h s_h \approx \frac{1}{L}\sum_{h=1}^{L} W_h s_h \quad (7)$$

This method can be implemented by starting with equidistant boundaries and iteratively refining them until the above condition is approximately satisfied.

---

[10] In computer simulation, this is exemplified by stratifying on basic block vectors (BBVs), where the BBV acts as $x$, an auxiliary variable correlated with CPI ($y$).

[11] This is the approach used in our study: we perform a large, costly random sample of the baseline configuration to obtain CPI ($y$) and related statistics for stratification, then reuse this stratification for future configurations. The CPI under another configuration is conceptually equivalent to $y$ at a different point in time.